\documentclass[man]{apa7}
\usepackage[american]{babel}
\usepackage{lineno}
\usepackage{amsfonts}
\usepackage{amsmath,amssymb}
\usepackage{bm}
\usepackage{bbm}
\usepackage{graphicx}
\usepackage{subcaption}
\usepackage{csquotes}
\usepackage[natbibapa]{apacite}  

\title{Merging public elementary schools to reduce racial/ethnic segregation}
\shorttitle{Merging public elementary schools}

\author{Madison Landry, Nabeel Gillani}
\affiliation{Plural Connections Group\\Northeastern University\\Boston Massachusetts, USA}

\authornote{To whom correspondence should be addressed:  landry.ma@northeastern.edu and n.gillani@northeastern.edu.}

\abstract{

Diverse schools can help address implicit biases and increase empathy, mutual respect, and reflective thought by fostering connections between students from different racial/ethnic, socioeconomic, and other backgrounds. Unfortunately, demographic segregation remains rampant in US public schools, despite over 70 years since the passing of federal legislation formally outlawing segregation by race. However, changing how students are assigned to schools can help foster more integrated learning environments. In this paper, we explore ``school mergers'' as one such under-explored, yet promising, student assignment policy change. School mergers involve merging the school attendance boundaries, or catchment areas, of schools and subsequently changing the grades each school offers. We develop an algorithm to simulate elementary school mergers across 200 large school districts serving 4.5 million elementary school students and find that pairing or tripling schools in this way could reduce racial/ethnic segregation by a median relative 20\%---and as much as nearly 60\% in some districts---while increasing driving times to schools by an average of a few minutes each way.  Districts with many interfaces between racially/ethnically-disparate neighborhoods tend to be prime candidates for mergers.  We also compare the expected results of school mergers to other typical integration policies, like redistricting, and find that different policies may be more or less suitable in different places. Finally, we make our results available through a public dashboard for policymakers and community members to explore further (\url{https://mergers.schooldiversity.org}). Together, our study offers new findings and tools to support integration policy-making across US public school districts.

}

\begin{document}
\maketitle

\section{Significance statement}
May 17th, 2024 marked the 70th anniversary of Brown vs. Board of Education, the landmark U.S. Supreme Court Case that outlawed racial segregation in public schools. Still, schools across America remain significantly segregated across racial/ethnic lines, which threatens to exacerbate achievement gaps and perpetuate inequalities in access to critical educational resources. Focusing on elementary schools, our study explores how, and how much, one particular policy---school mergers---might promote integration, and how much this strategy might increase travel times for families across different districts. Our findings suggest that across many districts, school mergers could meaningfully reduce segregation without imposing large travel burdens, revealing practical policy possibilities for fostering more equitable learning environments.

\section{Introduction}

\begin{figure}[h]
\centering
\hfill
\begin{subfigure}[b]{0.3\textwidth}
    \centering
    \includegraphics[width=\textwidth]{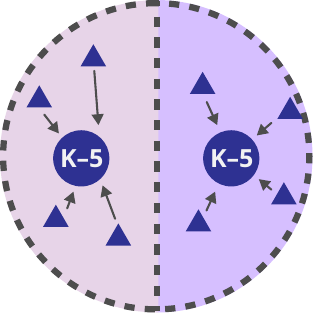}
    \caption{Before mergers}
    \label{fig:mergers_before}
\end{subfigure}
\hfill
\begin{subfigure}[b]{0.3\textwidth}
    \centering
    \includegraphics[width=\textwidth]{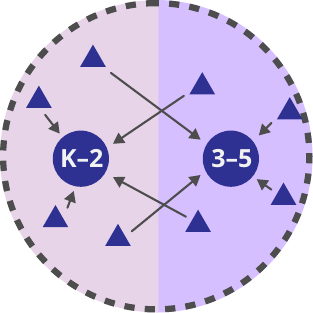}
    \caption{After mergers}
    \label{fig:mergers_after}
\end{subfigure}
\hfill
\begin{subfigure}[b]{0.3\textwidth}
    \centering
    \includegraphics[width=\textwidth]{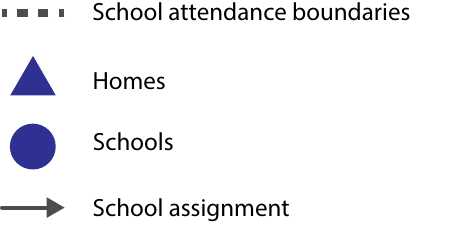}
\end{subfigure}
\hfill
\caption{\textit{School mergers} involve merging the attendance boundaries of adjacent schools and subsequently modifying the grades they serve to promote demographically-diverse classrooms. (a) Two adjacent K–5 schools within a district that happen to serve students who are demographically different from one another. (b) The schools can be merged so that one school serves only students in grades K–2 in the merged region, and the other serves only students in grades 3–5, thereby diversifying the student body of each school.}
\label{fig:mergers}
\end{figure}



There is substantial empirical support for the advantages that students experience when they interact with classmates from different races, ethnicities, \& socioeconomic backgrounds \citep{wells1994perpetuation, wells2016racially}. Some may argue that focusing on diversity diverts attention from addressing academic disparities. However, racial/ethnic, socioeconomic, and residential patterns contribute to educational segregation and widen achievement gaps \citep{reardon2019geography, fiel2018three}. Furthermore, the benefits of diversity extend far beyond academic achievement. 
Diverse educational settings can equip children with the skills needed to succeed in an increasingly interconnected world \citep{suarez2001globalization}. These settings create opportunities for engagement across different social groups, addressing stereotypical perceptions while fostering cognitive empathy and a genuine respect for diversity \citep{davies2011cross, salanga2019cognitive}.



Unfortunately, demographic segregation remains rampant in US public schools~\citep{gao2022segregation}, especially in elementary schools, which are smaller, greater in number, and draw from smaller catchment areas, increasing the link between residential and school segregation \citep{fiel2018three}.


Historically, school systems have explored a myriad of desegregation strategies, from the strategic placement of popular magnet programs, to ``redistricting'', or the redrawing of the school catchment areas that determine which students attend which schools~\citep{redrawing}. While geographic proximity (convenience) and residential segregation often dictate school enrollment, magnet \& choice schools aim to attract students from different demographics by offering specialized programs or curricula (e.g. STEM, arts, language immersion) \citep{ayscue2019magnets, hawkins2018texas}. While potentially effective in boosting diversity, there are challenges to maintaining diverse enrollments. Additionally, ensuring equitable access can limit the impact on reducing segregation, with affluent families often being in a better position to benefit from these programs~\citep{villarreal2010magnet}.  

A powerful ``default setting'' that influences how the majority of students across the US are assigned to public schools is school attendance boundaries. Attendance boundaries are the catchment areas that school districts draw to determine which neighborhoods are assigned to which schools. Prior work has demonstrated how ``redistricting'', or redrawing school district boundaries, might foster more diverse and integrated schools while, surprisingly, possibly also slightly reducing travel times~\citep{redrawing}. The study developed redistricting algorithms to simulate changes across nearly 100 US public school districts, building upon other literature in algorithmic redistricting---most notably, tools and methods developed to explore changes to voting districts in order to facilitate more equitable participation in the political process~\citep{becker2022redistricting, gurnee2021fairmandering}. 


Unfortunately, school redistricting is often hotly contested and opposed by community members.  Racial/ethnic integration is typically not a redistricting priority~\citep{castro2022richmond}---and when it is, it can surface concerns about splitting students up from their friend groups and other issues of ``community cohesion''~\citep{bridges2016eden}.  Some of these concerns may be coded reflections of parents' racialized preferences for schooling~\citep{billingham2016parents,hailey2021parents,hall2017migration}; others may reflect concerns about diminished access to ``quality'' schools, reflecting a deeper-seated perception of access to quality education as a ``zero-sum'' game~\citep{genevieve2018disintegration}. While some districts have experimented with voluntary integration plans~\citep{holley2009after}, many of which incorporate both boundary and choice-based policies, their impacts have been limited due to challenges in implementation and oversight. This suggests the importance of district-initiated and supported student assignment policy changes to reduce demographic segregation. Prior work suggests that nearly two-thirds of segregation can be attributed to the lines that are drawn \textit{between} school districts~\citep{fiel2013boundaries, barshay2024}, which fall under the purview of state legislatures. However, within-district boundaries---like school attendance boundaries, which districts themselves have jurisdiction over redrawing---still contribute to segregation in a nontrivial way, and so, may serve as powerful and more practical levers for reducing demographic segregation across schools. 

The important role school districts can play in integrating schools is further underscored by renewed efforts by the US Department of Education to support districts in exploring and implementing new integration strategies---for example, through their 2023 ``Fostering Diverse Schools'' grant program~\citep{jacobson2023fostering}. Such efforts stand to benefit from empirical research into different types of integration strategies that may help reduce segregation and foster more equitable learning environments across districts. 



In this study, we investigate one particular integration strategy that districts might explore: ``school mergers'', illustrated in Figure \ref{fig:mergers}. School mergers involve merging the attendance boundaries of pairs or triples of adjacent schools and subsequently modifying the grades they serve in order to promote demographically diverse classrooms. For example, consider two adjacent K–5 schools within a district (see Fig. \ref{fig:mergers_before}). Despite their vicinity, such schools may serve students who are demographically very different from one another. However, the attendance boundaries of these schools might be \textit{merged} (in this case, paired) so that one school serves only students in grades K–2 in the merged region and the other serves only students in grades 3–5 (Fig. \ref{fig:mergers_after}). In this way, students from different backgrounds are more likely to encounter, and possibly befriend, those who are different from them---and have the opportunity to deepen these connections as they progress to higher grades. Such an integration strategy may also be preferable to others (most notably, redistricting), because mergers do not split friend groups within grade cohorts.

School pairings and triplings were common in the era of court-ordered desegregation following Brown v. Board~\citep{welch1987new}. In recent years, district leaders have explored them in places like Charlotte, NC~\citep{helms2020charlotte}, Richmond, VA~\citep{arriaza2019richmond}, and Washington, D.C.~\citep{lumpkin2024dc}. Despite some of their appealing properties, in some cities, they have failed to pass due to community backlash resembling the types that often surface in redistricting initiatives~\citep{arriaza2019richmond,swanson1965mergers}. In other cities, they have passed, but raised questions about longer-term sustainability and potential for positive academic impacts~\citep{helms2020charlotte}. Finally, with mergers and other integration policies, there remains the perennial concern of ``White flight'', or the departure of White/affluent families from their assigned schools following re-assignment, impeding progress towards more integrated environments~\citep{reber2005flight,nielsen2020denmark,macartney2018boards}.  

For these reasons, and the broader sociopolitical complexities of integration, mergers---just like any other policy strategy---are unlikely to address segregation on their own. However, there continues to be limited empirical evidence on how much different strategies, including school mergers, \textit{might} reduce segregation in different places; what the ``costs'' (particularly in terms of times) of such policies might be; and how the expected impacts of such policies might compare to other integration policies in the same place. Offering this empirical evidence does not obviate the sociopolitical frictions involved in desegregation policymaking, but may help policymakers and community members gain a clearer understanding of what might be possible in their own communities. \textit{The purpose of this study is to help paint this picture of what might be possible}. Specifically, we ask: how much might we expect elementary school mergers to reduce racial/ethnic segregation across large US public school districts, and what costs (for example, in terms of increases in travel times to school) might these reductions come at? 

We develop an algorithm for exploring school pairings and triplings and simulate elementary school mergers across the 200 US school districts with the most elementary school students enrolled in closed-enrollment elementary schools (i.e., schools whose attendance is determined by those living within the school's attendance boundary). Collectively, these districts serve nearly 4.5 million students and tend to be more segregated, and have more students of color, than the broader set of districts across the US (see Materials and Methods for more details on the sample of districts). 

\begin{figure}
\centering
\includegraphics[width=.7\linewidth]{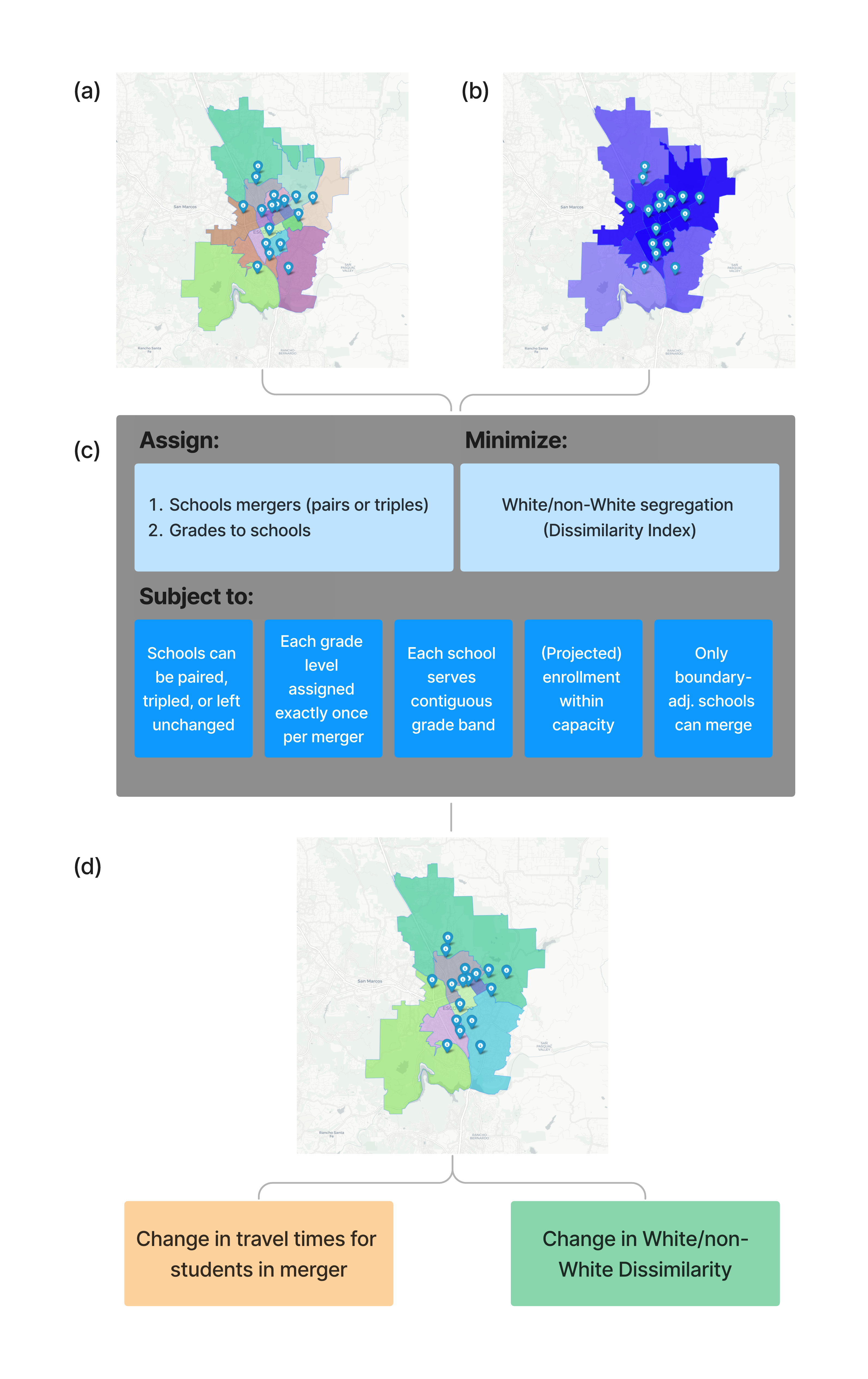}
\caption{\small{Diagram outlining the basic algorithm for school mergers, showcasing Escondido Union school district (NCES ID: 0612880) in California as an example.  Plot (a) shows status quo elementary school attendance boundaries; (b) illustrates the demographic distribution of elementary students in the district (darker blue indicates higher concentration of students of color); (c) outlines the algorithm's key decision (assignment) variables, objective function, and constraints; (d) specifies the algorithm's key outputs: a map of merged school attendance boundaries, and expected impacts on segregation in the district, and expected impacts on students' travel times.}}
\label{fig:opt_model}
\end{figure}

We simulate school pairings/triplings across the closed-enrollment schools in these districts that seek to minimize segregation---measured by the dissimilarity index~\citep{massey1988dimensions}---between White students and students of color (i.e., students classified as Hispanic/Latinx or some other racial category except for White). The inputs into the simulation include student enrollments per racial/ethnic group, per grade, per school, as well as a number of constraints that the simulation must adhere to while searching for possible mergers that could reduce segregation. These constraints influence which schools can be merged with which others; which grades each merged school can serve; and other factors (see Figure~\ref{fig:opt_model}). Our results reveal an expected 20\% relative decrease in median segregation, which we estimate would require a median 3.7-minute increase in driving times each way for students.  Further analyses reveal that patterns of racial/ethnic spatial autocorrelation are positively correlated with how much mergers might reduce segregation in different districts---and that different districts may benefit more or less from different integration policies (i.e., mergers or redistricting). These findings, which we make available to school districts and community members through a public dashboard (\url{https://mergers.schooldiversity.org/}), offer practical insights that could help inform desegregation policy-making across US public schools.

\section{Results}

\begin{figure*}[h]
\centering
\hfill
\begin{subfigure}[b]{0.45\textwidth}
    \centering
    \includegraphics[width=\textwidth]{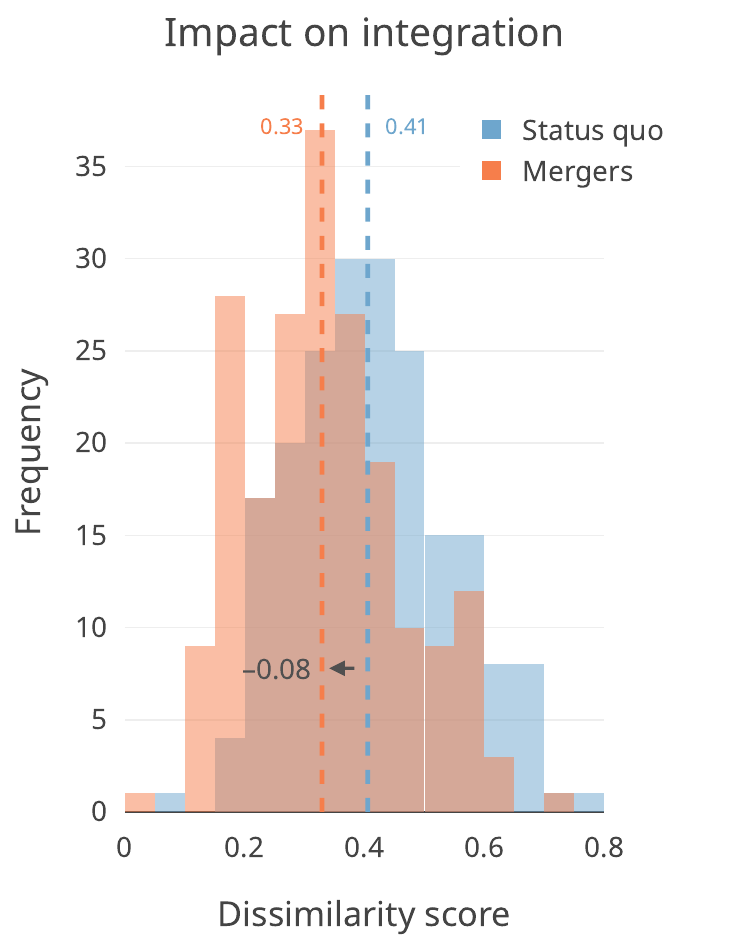}
    \caption{Dissimilarity}
    \label{fig:dissimilarity_histogram}
\end{subfigure}
\hfill
\begin{subfigure}[b]{0.45\textwidth}
    \centering
    \includegraphics[width=\textwidth]{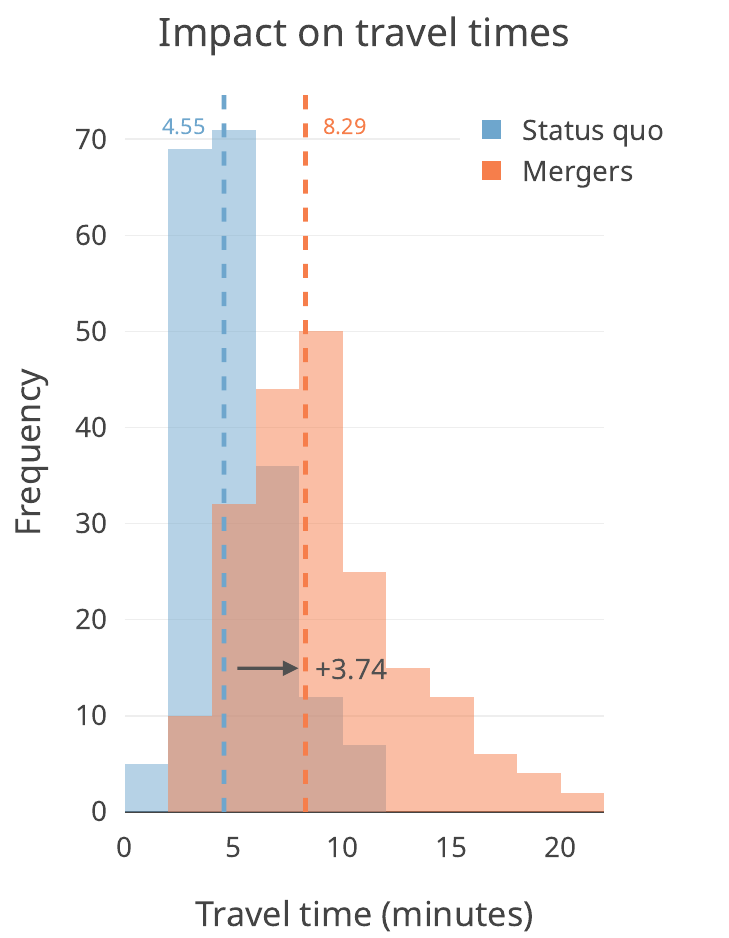}
    \caption{Travel times for students who switched}
    \label{fig:travel_times_histogram}
\end{subfigure}
\hfill
\caption{Summary histogram plots for the top 200 districts by population size. Results show that merging (pairing or tripling) schools might increase median travel time for students who would be switching schools from 4.5 minutes to 8.3 minutes, and decrease median segregation (dissimilarity) across the depicted districts from 0.41 to 0.33 ($-20.\%$).}
\label{fig:overall_histograms}
\end{figure*}

We measure segregation using the dissimilarity index~\citep{massey1988dimensions}, where the two groups of students correspond to White students and students of color. A score of 1 indicates perfect segregation (all White students and students of color are concentrated entirely in separate schools), and a score of 0 indicates perfect integration (i.e., each school's racial/ethnic demographics reflect district-level proportions). While the dissimilarity index suffers from several shortcomings~\citep{james1985seg,winship1978dissim}, prior work on algorithmic redistricting for school integration has revealed similar results when optimizing for dissimilarity and other measures of segregation, like the normalized exposure index~\citep{redrawing}.

Our algorithm seeks to identify pairings and triplings that might minimize this dissimilarity measure while respecting several constraints depicted in Figure~\ref{fig:opt_model} and described in more detail in the Materials and Methods section. One notable constraint is on school capacities: mergers cannot occur if any of the schools in the merger would subsequently exceed their capacity. Unfortunately, school capacities are not reported by the Department of Education; therefore, we use the maximum enrollment observed at the school over the past six years (from 2016/2017 to 2021/2022) as a proxy. This is an imperfect measure that does not account for schools that are chronically over or under-enrolled; therefore, an important caveat is that our simulations may, at times, suggest inviable mergers, or fail to propose viable ones, due to this estimation procedure.

After identifying which schools should be merged and subsequently which grade levels each school should serve, the algorithm produces a number of outputs describing various expected family and district-level impacts of the new student assignment regime. These outputs---like expected impacts on travel times, school demographics, and other factors---are perhaps most easily explored interactively and visually through the dashboard we release with this project: \url{https://mergers.schooldiversity.org/}. 

\begin{figure}
\centering
\includegraphics[width=\linewidth]{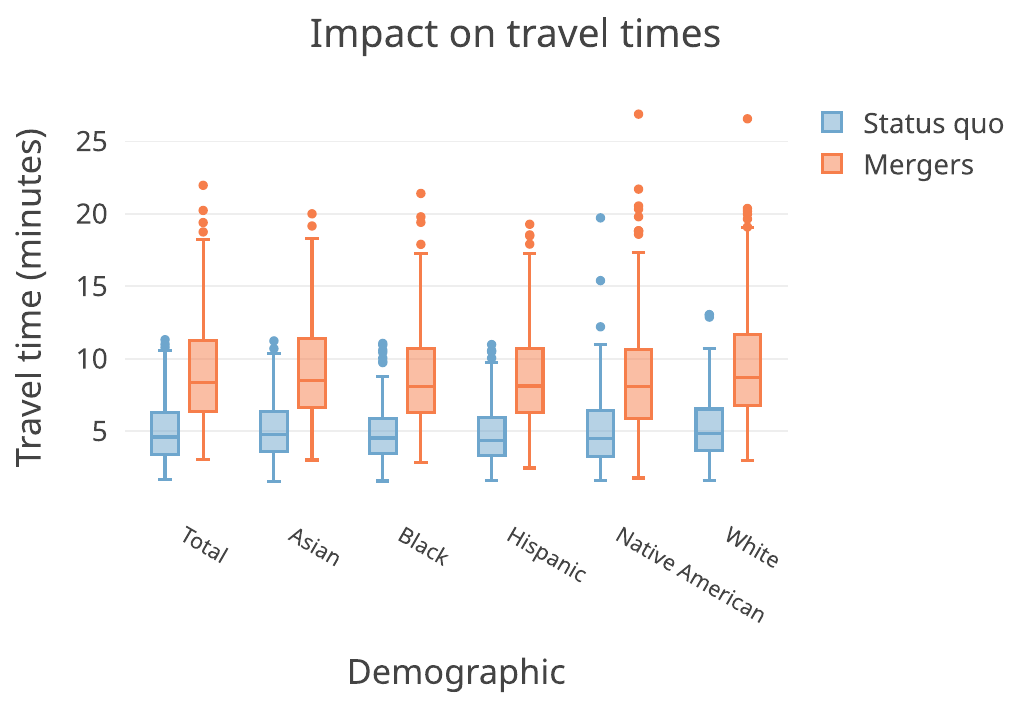}
\caption{Expected increases in average travel times for students from different demographic groups who would be involved in a school merger. Differences across demographics are not statistically significant.}
\label{fig:travel_times_comprehensive}
\end{figure}

We summarize  expected decreases in dissimilarity and changes in family travel (driving) times to schools in a hypothetical scenario where the districts in our sample implement our algorithmically-suggested school pairings and triplings. As Figure \ref{fig:overall_histograms} reveals, simulating elementary school pairings and triplings across these districts shows an expected median 20.\% relative decrease in segregation (median dissimilarity score change of 0.41 to 0.33), which could require a median 3.7-minute increase in travel times each way for students who would be switching schools (i.e., from 4.5 to 8.3 minutes) and require a median of 36\% of students to switch schools.  Figure \ref{fig:travel_times_comprehensive} displays expected changes in travel times for students who would be involved in a merger, across demographics. Differences between different demographic groups appear to be small, suggesting---importantly---that no single group appears disproportionately affected by travel time increases, contrary to historical integration strategies like long-range ``busing''~\citep{delmont2016busing}. While we estimate driving time increases to generally be small, estimating travel time changes for other modalities (e.g. bus routes) would require additional modeling, a la~\citep{bertsimas2019optimizing}. We also note that for those presently in walking distance to their elementary schools, pairing/tripling schools would likely reduce walkability---and therefore, may serve as a key point of opposition for families.

\begin{figure}
\centering
\includegraphics[width=\linewidth]{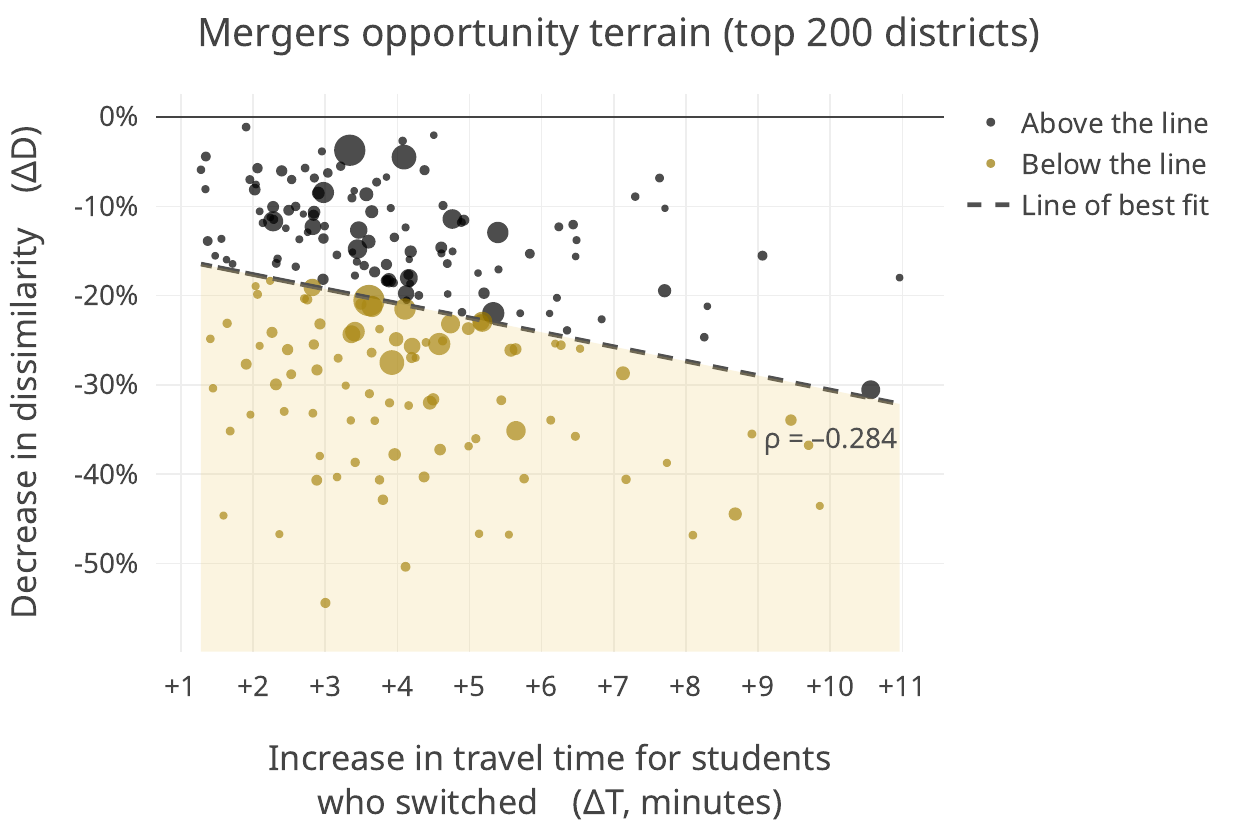}
\caption{Scatter plot and OLS line of best fit for change in dissimilarity ($\Delta D$)  over change in travel time ($\Delta T$) for students who would switch schools under elementary school mergers, together representing the trade-offs \& opportunities for school districts in using school merging as an integration strategy. Marker size is scaled to be proportional to the district's population that attends closed-enrollment elementary schools. A Spearman rank correlation coefficient of $\rho = -0.284$ ($p<0.0001$), serves as a reference for which districts demonstrate the potential for ``integration arbitrage'': achieving relatively higher levels of integration at a lower (travel time) cost. Districts below the threshold have a greater decrease in dissimilarity score per increase in travel time than districts below the threshold have. The plot illustrates that, for some districts, a large decrease in dissimilarity score is traded for a small increase in travel time, suggesting that the mergers approach is a worthwhile approach for some (but likely not all) districts.}
\label{fig:opportunity_plot}
\end{figure}

Figure \ref{fig:opportunity_plot} visualizes, for each district, how much decreases in dissimilarity might ``cost'' in terms of increases in travel times for families. We use ordinary least squares (OLS) regression to explore this relationship and contextualize the results for each district in relation to others in the sample. The plot reveals a Spearman rank correlation coefficient of $\rho = -0.284$ between these values across districts---indicating that, as expected, larger decreases in segregation induced by mergers are typically accompanied by larger increases in travel times.  However, this relationship is not as strong as we might expect---suggesting the possibility for ``integration arbitrage'' (i.e., achieving large amounts of desegregation at a small travel cost) in certain districts. 

Of course, the relatively small median increase in driving times observed across our sample after simulating mergers likely stems in part from the adjacency constraint in the algorithm: only schools with adjacent attendance boundaries can be merged. Still, as Figure~\ref{fig:travel_times_histogram} shows, many districts could still experience large increases in driving times following a merger---suggesting that the adjacency constraint alone does not guarantee minimal driving time increases. This could be due to a number of factors like geographically-expansive boundaries; populations that are concentrated at the extreme corners of boundaries (and far away from schools in other boundaries); and/or existing road networks that do not permit efficient travel to newly-assigned schools, among others.

\subsection{Sensitivity analyses}\label{sec:sensitivity}
We conduct four sensitivity analyses. The first explores how results might differ under an alternative objective measure of segregation---Black/Hispanic vs. White/Asian segregation---and how that compares to the White/students of color segregation we analyze throughout the paper. The second explores how impacts on integration might differ if some subset of families involved in a school merger opted out of their schools. The third investigates how results might differ under different minimum school capacity constraints. Finally, the fourth explores results in a setting where \textit{between} district mergers are also allowed. Details are available in the Supplementary Materials.

\subsubsection{Optimizing for Black/Hispanic and White/Asian integration}

The application of school mergers as an integration strategy requires factoring in local context. Different districts have unique demographic distributions, geographies, or other characteristics that may be pertinent. While the main results minimize dissimilarity between White students and students of color, in some districts, a salient dimension of segregation may be segregation between Black and Hispanic students and their White and Asian counterparts---though this, too, is a contentious dichotomy~\cite{orfield1994asian}. We therefore conduct additional simulations that seek to minimize Black/Hispanic and White/Asian dissimilarity across districts and find nearly identical results. In some districts, however---particularly those with a large concentration of Asian students---there are notable differences in the relative expected impacts of optimizing for either notion of dissimilarity. See the Supplementary Materials for more details.

\subsubsection{Opt-out analysis}

We follow the method described in~\citep{redrawing} to estimate how the prevalence of magnet and charter schools might impact the extent to which families remain in their new schools post-merger. In particular, for each school district in our sample, we identify which charter and magnet schools fall within the boundaries of that district and the ratio of enrollment, by race/ethnicity, in the ``choice'' options relative to closed-enrollment district options (for example, if 100 White students attend the closed-enrollment elementary schools in the district, and 20 attend charter or magnet schools, this ratio would be 0.2). We then simplistically model a scenario where students attending a school that our algorithm pairs or triples with another school opt out at a rate indicated by their demographic-specific ratio. Doing this reveals a median expected decrease in dissimilarity of approximately 17\%, down from the 20\% in our main results. Of course, the impacts could be more severe if families opt for private options or move out of the district entirely. Nevertheless, these results suggest that pairings and triplings can still meaningfully advance desegregation efforts even if some families do not adhere to the new student assignment policies.

\subsubsection{Changing the minimum school enrollment threshold}

Our main results require that a school's enrollment post-merger must be at least 80\% of its pre-merger enrollment. This is because school funding is often tied to enrollment numbers, and therefore, allowing drastic reductions in enrollment could lead to large funding cuts, as well as other unfavorable impacts like decreased parental involvement or extracurricular participation. We find that loosening the constraint on minimum school capacity yields more schools involved in mergers, and larger decreases in dissimilarity.

In particular, setting the minimum required enrollment to 0\% allows the possibility of schools that serve no students post-merger---i.e., a ``school closure''. School closures, while sometimes necessary due to declining student enrollments and other demographic forces, can create disruptive  ``closure effects'' such as reduced morale and drop in academic performance for the students they impact~\citep{kirshner2010tracing}. Nevertheless, some districts may find closures desirable, for saving costs on under-utilized or aging schools, or as an opportunity for renovation/demolition. School closures have become a particularly relevant topic post-COVID, as many districts are facing ``fiscal cliffs''---and accordingly, difficult resourcing decisions---with the imminent end of pandemic-era funding~\citep{roza2024closures}. 

We set the minimum enrollment requirement to 0\% to allow for the possibility of closures and simulate school mergers. Results suggest pairing or tripling 7,116 across our 200 districts, but only closing 24 of them (i.e., 24 would no longer serve any students post-merger)---though 312 would serve 50\% or less of their original enrollment post-merger. These results suggest that in some districts, school closures might advance desegregation objectives, but for the reasons above, policymakers should proceed with caution as they consider them. See Supplementary Materials for more details.

\subsubsection{Allowing inter-district mergers}\label{sec:sensitivity_interdistrict}

Our primary simulations explore within-district mergers, even though a majority of school segregation can be attributed to how the lines \textit{between} districts themselves are drawn~\citep{fiel2013boundaries,barshay2024}. In practice, between-district mergers, like other inter-district integration policies~\citep{potter2020integration}, may be difficult to enact because they require coordination between multiple local educational agencies. However, they may offer pathways to integration that are not possible solely within certain districts~\citep{simko2023seg}.

We conduct simulations that allow a district's schools to merge either with other adjacent schools in the same district, or with adjacent schools in a neighboring district. Simulating these inter-district pairings and triplings yields an expected decrease in dissimilarity of just over 12\% and would require students involved in a merger to travel, on average, approximately 4.5 minutes longer to school each way. 
 The longer travel times make sense given the larger geography that inter-district mergers would operate over. The relatively smaller decrease in segregation (compared to the main intra-district results presented above) also makes sense given that inter-district mergers are only possible between schools on the borders of the districts; in the event that schools further from the borders also contribute to between-district segregation, which is likely, only allowing mergers between border schools will limit how much integration between districts is possible.

\subsection{Case studies}\label{sec:case_studies}

To make our primary results more concrete, we explore two specific districts as case studies: one demonstrating a comparatively smaller potential decrease in segregation due to mergers, and another, a larger potential decrease.


\subsubsection{Miami-Dade}

Miami-Dade is the largest district in our sample, with an elementary school population of 118,731 elementary students across 211 closed-enrollment schools. The majority of these students are classified as Hispanic/Latinx, with White students making up 7\% of the population. The district has a dissimilarity index of 0.58---higher than the median of 0.41 across the full sample of districts---suggesting more opportunity for mergers to foster less segregated schools.

Simulating mergers proposes pairing 118 schools into 59 clusters and tripling 24 schools into 8 clusters, yielding a resulting dissimilarity index of 0.56, a relative decrease of just 3.7\%. These pairings and triplings would involve 36.9\% of elementary school students in the district (approximately 43,773), and hence, require them---at some point---to switch elementary schools. For these students, we estimate that their average travel time to school (each way) would increase from 4 minutes to 7 minutes. Despite the potential for integration, mergers appear to offer limited gains in integration for this district.

These results may be due to an interesting mix of social and geographical factors.
As illustrated in Fig. \ref{fig:miami-dade}, the schools along the coast comprise relatively more White students than inland districts. Since these coastal schools are surrounded by water on at least one side but as many as all sides, there is minimal interface between clusters of White students and clusters of students of color in the district. Ultimately, since the underlying mechanism of school mergers involves pairing or tripling adjacent schools, Miami-Dade serves as an example of a school district with lower potential for integration using this strategy. Districts like Miami-Dade may wish to explore non-adjacent mergers, though, given the greater impact such mergers are likely to have on driving times, and the fact that families often prefer contiguous attendance areas~\cite{mcps2021analysis}, these policies may prove more difficult to pass in practice. Multi-group measures of segregation like~\cite{elbers2021method} may also help highlight dimensions of segregation that may be masked by a dichotomous measure like the dissimilarity index.


\subsubsection{Plano Independent School District (ISD)}

Plano Independent School District has approximately 17,553 students across 43 closed-enrollment elementary schools, with the majority classified as White or Hispanic/Latinx students and White students making up 34\% of the population. The existing dissimilarity index for Plano ISD is 0.32, below the median value of 0.41.

Simulating school mergers yields a pairing of 36 elementary schools into 18 clusters and tripling 3 schools into a single cluster. These changes would yield a dissimilarity index of 0.15, a relative decrease of 54\%. Remarkably, pairing/tripling schools could \textit{halve} existing levels of racial/ethnic segregation across the district's elementary schools. In this simulation, nearly half of the elementary-aged students (48.0\%, approximately 8,430) would be required to switch schools at some point. For these students, we estimate that their average driving times would increase by approximately 3 minutes each way (from 4 to 7 minutes).

\begin{figure*}
\centering
\hfill
\begin{subfigure}[b]{0.45\textwidth}
    \centering
    \includegraphics[width=0.7\textwidth]{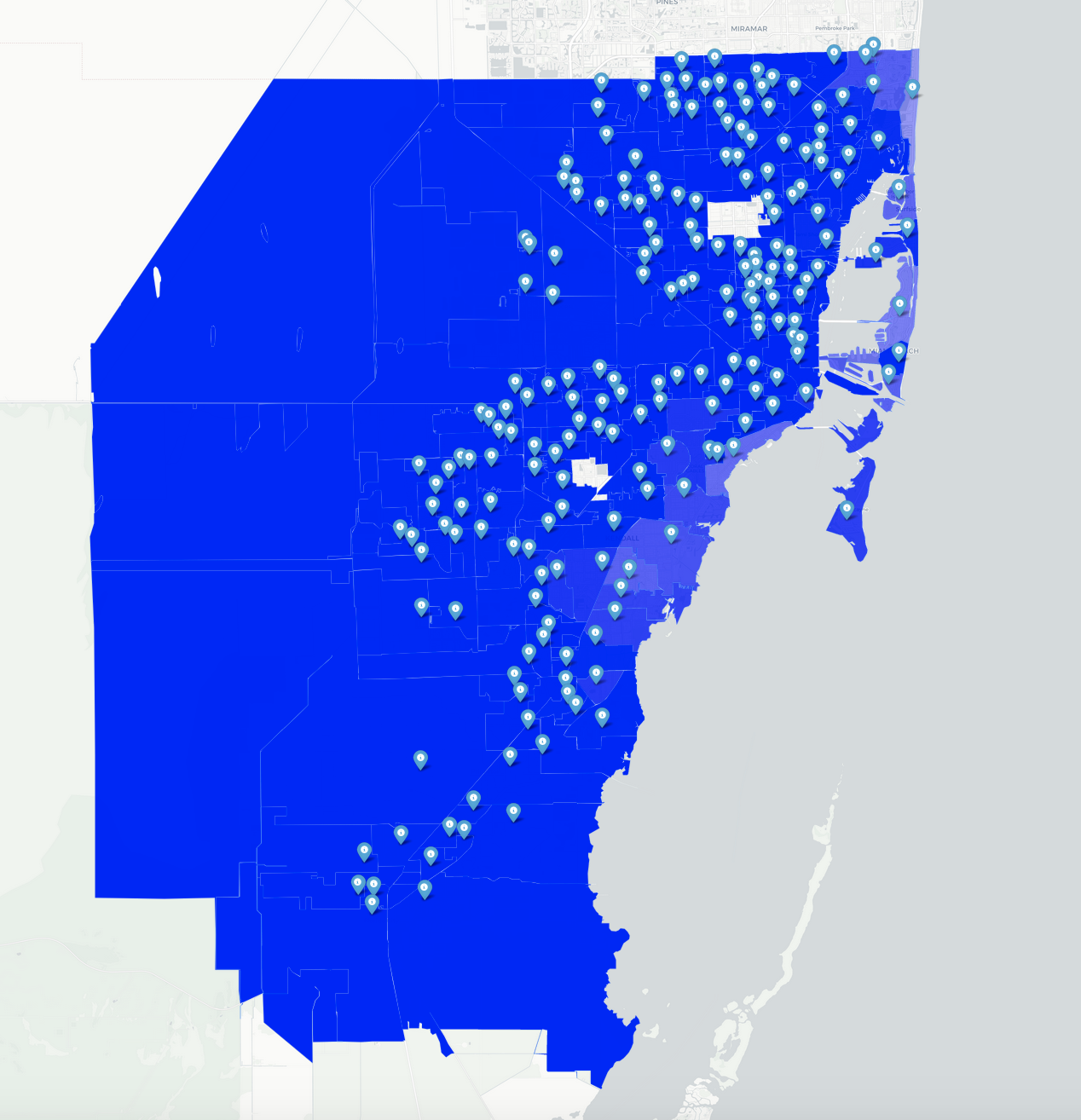}
    \caption{Miami-Dade \\ (NCES ID: 1200390)}
    \label{fig:miami-dade}
\end{subfigure}
\hfill
\begin{subfigure}[b]{0.45\textwidth}
    \centering
    \includegraphics[width=0.9\textwidth]{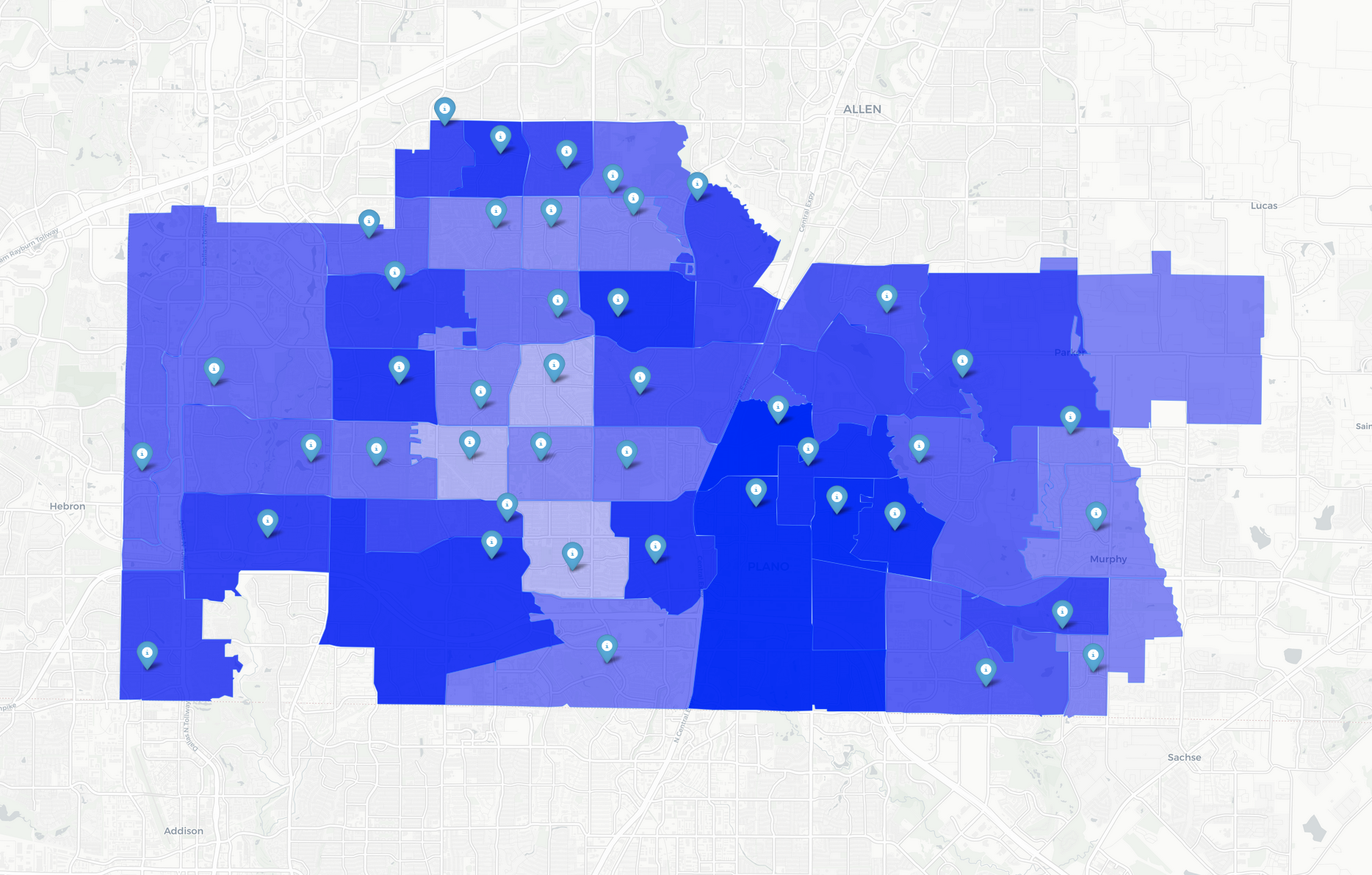}
    \caption{Plano ISD \\ (NCES ID: 4835100)}
    \label{fig:plano}
\end{subfigure}
\hfill
\vfill
\begin{subfigure}[b]{\textwidth}
    \centering
    \includegraphics[width=0.7\textwidth]{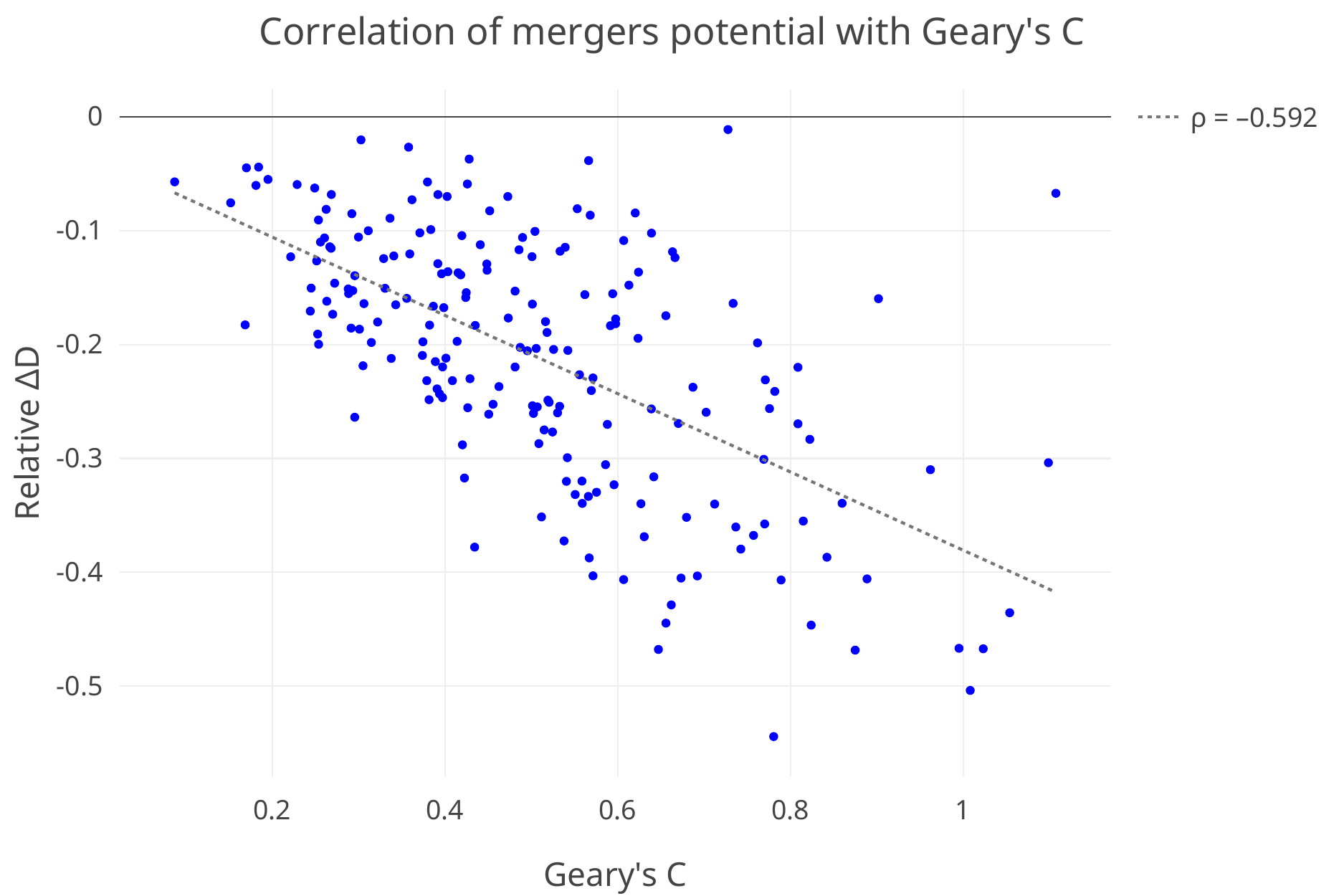}
    \caption{Correlation with Geary's $C$}
    \label{fig:gearys_c}
\end{subfigure}
\caption{\small{School mergers are less likely to foster meaningful integration in districts with few interfaces between attendance zones with different demographics and more likely in districts with more such interfaces ((a) and (b), respectively). In these plots, darker blue indicates higher concentration of elementary students of color.  (c) Scatter plot and OLS line of best fit ($\Tilde{\Delta D} = -0.344 C - 0.037$; $r=-0.611$, $p<0.0001$) for the change in dissimilarity ($\Delta D$) over Geary's $C$, for the 200 districts in our sample. A Spearman rank correlation coefficient of $\rho = -0.592$ suggests a moderate, negative correlation, where greater values of $C$---representing stronger spatial anticorrelation---indicate higher potential for district integration via school mergers.}}
\label{fig:race_maps}
\end{figure*}

In contrast to Miami-Dade, Plano ISD has a greater number of interfaces between areas of majority color and majority White areas, akin to a checkerboard pattern (Fig. \ref{fig:plano}). This reflects the propensity for even diverse cities to reflect segregation at smaller scales, as communities often self-select into homogenous neighborhoods and communities~\citep{nilforoshan2023seg}. Since school mergers rely upon a mechanism of pairing adjacent schools to adjust student demographics, the presence of diverse populations in adjacent attendance boundaries makes this strategy potentially highly effective in this district.

\subsection{Spatial autocorrelation as a heuristic}

The two case studies (\S\ref{sec:case_studies}) suggest that the degree to which school pairings and triplings might reduce dissimilarity may be associated with the extent to which a district exhibits demographic spatial clustering.

Geary's $C$ is a statistical measure of spatial autocorrelation (i.e. clustering). In particular, when $C$ is large, it suggests that geographic units within a district (i.e., attendance boundaries) with different proportions of students from the same demographic groups tend to be adjacent. Accordingly, a low value of $C$ signifies high racial segregation across attendance boundaries and limited geographic interfaces between diverse populations throughout the district. For instance, Geary's $C$ for Miami-Dade (Fig. \ref{fig:miami-dade}) is $0.428$, whereas for Plano ISD (Fig. \ref{fig:plano}), it is $0.781$.

We compute $C$ for all districts in our sample and compare it to the expected decrease in segregation observed for those districts after simulating school mergers, illustrated in Figure \ref{fig:gearys_c}. A Spearman rank correlation coefficient of $\rho = -0.592$ suggests a strong negative association between both measures---i.e., districts with strong patterns of demographic spatial autocorrelation also tend to be those where pairings/triplings yield the largest decreases in dissimilarity.  However, the correlation is far from perfect, and outliers like Birmingham City---where $C = 1.1$ but $\Delta D=-0.067$---illustrate the important role other locale-specific factors (like school capacities) might play in enabling large gains in integration from school mergers. Still, Geary's $C$ can serve as a useful visual heuristic that districts may use to quickly assess the potential for integration across different student or school characteristics.

\subsection{Comparison to redistricting}

School mergers are just one of many integration strategies that districts might pursue. Different strategies are likely to work with varying degrees of effectiveness in different places, depending on each place's unique demographics, geographies, and other characteristics. To help school districts better understand which integration strategies might be more or less effective in their districts, we compare our school merger simulation results to those from another integration strategy: the redrawing of school attendance boundaries, or ``redistricting''~\citep{redrawing}. Specifically, for each district in our sample, we explore how the potential gains in integration vs. costs in travel times that school mergers might induce compare to the gains in integration vs. costs in number of students who would be required to switch schools under a given redistricting plan. For redistricting, we focus on school switching instead of travel times as the primary ``cost'' of integration given results from~\citep{redrawing}, which showed that in many cases, travel times might actually \textit{decrease} under integration-promoting redistricting plans.

\begin{figure}
\centering
\includegraphics[width=.95\linewidth]{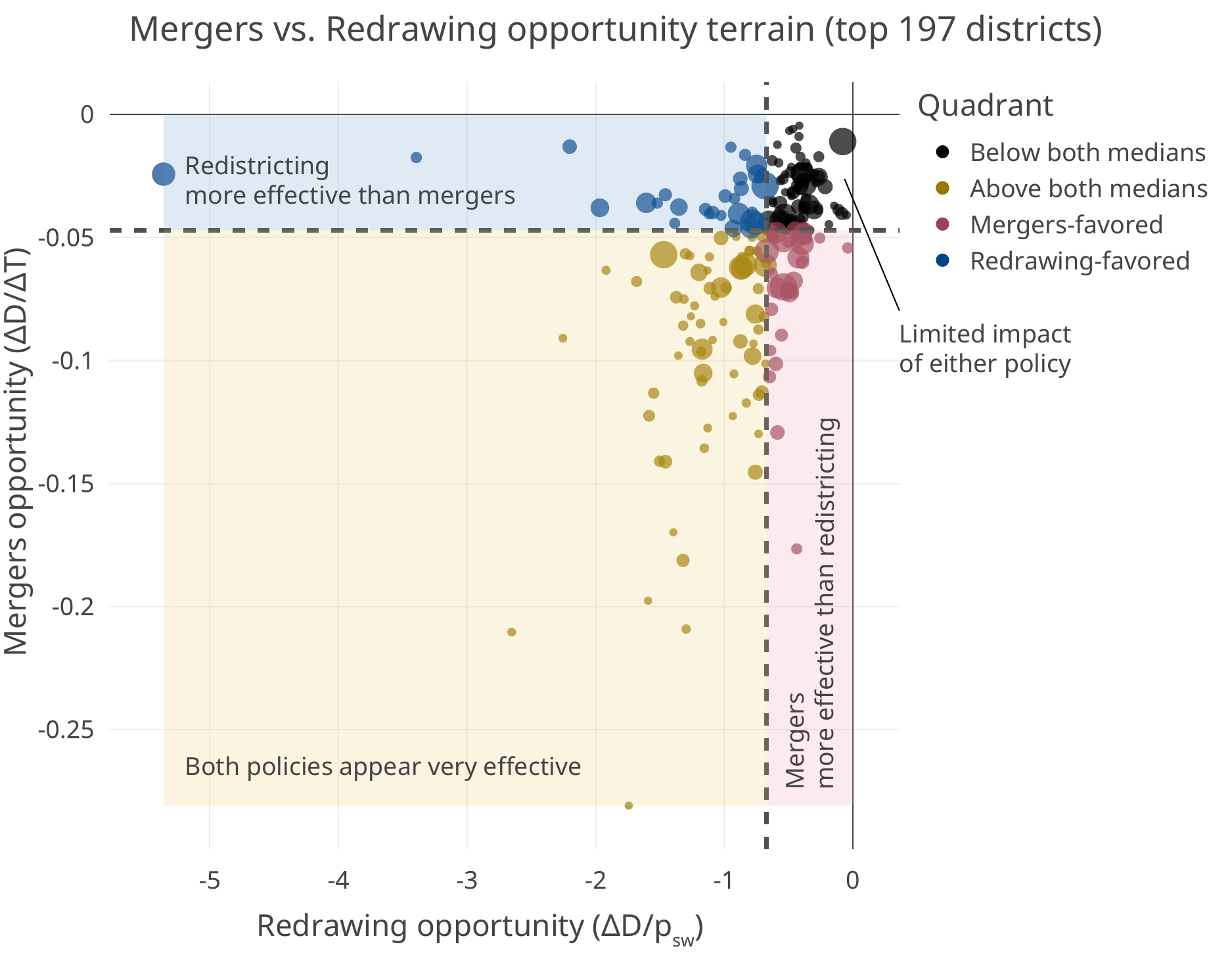}
\caption{Scatter plot cross-comparing the effectiveness of the school mergers integration strategy (vertical) with the redrawing school attendance boundaries integration strategy (\citep{redrawing}, horizontal). The y-axis represents the trade-off in relative change in dissimilarity index versus absolute change in travel time for the mergers strategy ($\Delta D/\Delta T$). The x-axis represents the trade-off in relative change in dissimilarity index versus the proportion of students who would switch schools for the redistricting strategy ($\Delta D/p_{\mathrm{sw}}$). In both cases, a more negative X and Y value is generally more favorable, representing a higher decrease in segregation at lower increase in travel times or students being rezoned, respectively.}
\label{fig:crossover_plot}
\end{figure}

Figure \ref{fig:crossover_plot} illustrates the trade-offs of each strategy in the form of a scatter plot with four quadrants, for the top 197 school districts by population size (redistricting results are drawn from~\citep{redrawing} and~\citep{gillani2023las}; three districts in our sample of 200 did not have corresponding redistricting results, likely due to performance bottlenecks that prevented their successful simulation). The figure shows a moderate correlation ($\rho=0.46$, $p<0.0001$) between both integration strategies, suggesting that, in general, districts where mergers offer a high-yield, low-cost approach to integration also tend to be districts where redistricting offers the same (and vice versa). Crucially, Figure~\ref{fig:crossover_plot} also reveals that there are different types of districts---some where both mergers and redistricting may yield large integration gains at comparatively lower costs; some where neither strategy produces large gains in integration relative to costs; and some where one policy appears to yield greater gains relative to costs compared to others. A thorough exploration of these results---including the specific characteristics of districts that make them more or less suited for either strategy---is beyond the scope of this paper. We include this data in our data release so interested readers may explore these differences further. Importantly, these results highlight how the simulation of different integration strategies might help districts determine which ones offer more or less promise in their districts, and therefore, could serve as a useful resource for local policy-makers.

\section{Discussion}

Our findings highlight that, across many districts, school mergers can serve as an effective school integration strategy without imposing large driving time increases for families. Interestingly, several districts exhibit a potential for ``integration arbitrage'', or large decreases in segregation that may be achieved through comparatively small driving time increases. One factor associated with the extent to which mergers might foster more integration in a school district is the strength of spatial autocorrelation in the district's demographics: districts with more interfaces between racially/ethnically diverse elementary attendance boundaries are more likely to benefit from pairings and triplings as an integration strategy. Finally, when comparing school mergers to another common integration strategy---redistricting---we find that in some districts, either strategy may yield promising reductions in segregation without imposing large travel time or student-school-switching costs; in others, neither strategy appears to be particularly effective; and still in others, one strategy may be more effective than the other. Importantly, the findings we present for the 200 districts in our sample of study may not generalize to other districts around the US. To enable researchers, practitioners, and community members to explore how mergers might unfold and what impacts they could have in a broader sample of districts, we release an interactive public dashboard (\url{https://mergers.schooldiversity.org}). 

We note several limitations in our preliminary study, which we believe can inspire future work on this nascent but important topic. For one, we found little literature on the longer-term academic, socioemotional, and other impacts of school pairings and triplings. While a long line of studies like~\citep{johnson2011desegregation} highlight the positive impacts of desegregation efforts more generally, an open question remains around whether and how requiring additional school transitions at the elementary ages might impact these outcomes. Prior literature has highlighted the potentially disruptive effects of elementary-to-middle school transitions~\citep{brookingsGrades}. However, these transitions are often accompanied by changes to friendship networks, including split friendship groups in cases where an elementary school might split-feed into more than one middle school. School pairings and triplings keep within-grade friend cohorts together, perhaps mitigating these disruptive effects---though additional research is required to fully understand their impacts. 

Another class of limitations includes the potential inconveniences that mergers might impose on families. As described in the results section, mergers may reduce walkability to schools---an issue that may be mitigated by imposing travel or distance constraints in the mergers algorithm to only allow mergers that are likely to preserve walkability (or other commute-related desiderata). Mergers may also split siblings of similar ages up into different schools---posing logistical or other challenges to parents looking to participate in their children's school communities (though this may happen at later ages anyways, e.g., when the older child(ren) transition from elementary to middle school). Fortunately, there are examples of case studies like~\citep{sundaram2024merger} highlighting how mergers, when conducted with thoughtful foresight and planning, can have positive impacts on students and families. Learning from such case studies before implementing mergers in other places might help mitigate such inconveniences.

Our use of the dissimilarity index to measure segregation between White students and students of color in school districts poses another limitation. We recognize that this dichotomous measure risks essentializing racial differences as well as disregarding the diversity within the broader `students of color' category. However, given the persistent disparities in educational access \& outcomes along these lines, and associations between race/ethnicity and other measures of segregation that often accompany achievement gaps (like socioeconomic status-based segregation~\citep{reardon2019geography}), it remains a salient lens for identifying patterns of segregation and informing efforts to create more integrated \& equitable schools. Fortunately, it is relatively straightforward to simulate school mergers across other student groups and definitions of diversity. Future work may consider less traditional, yet still important, definitions of diversity---like diversity in place of origin, neurodivergence, disability status, school curriculum/program offerings, and even typical academic performance measures. We release the data and code from this paper to support other researchers in these explorations.



There are also several simplifications and assumptions we make in order to produce preliminary results. Perhaps most notably, we are limited in the extent to which we can anticipate how families might respond to mergers by opting out of assigned schools. Our sensitivity analysis suggests that even under increased opt-out rates, mergers are still likely to meaningfully contribute to district-level integration objectives. Still, prior case studies like~\citep{arriaza2019richmond,swanson1965mergers} suggest that parents sometimes have strong reactions to the possibility of school pairings and triplings. This means that school opt-outs are likely to continue to threaten progress towards integration---though, perhaps, not entirely undo it. Indeed, a recent pairing in Charlotte did not witness disproportionate levels of flight post-pairing~\citep{helms2020charlotte}. Reducing flight and therefore securing the integrative impacts of student assignment policy changes will likely require a thoughtful approach to community engagement, among other efforts~\citep{gillani2023air}. 


Finally, additional practical considerations include 1) how districts might account for increased transportation costs that mergers might introduce, and 2) whether or not the physical infrastructure of elementary schools may be flexible enough to accommodate/focus on different grade cohorts. Working closely with school districts to obtain new datasets on school facilities, and exploring algorithms that try to jointly optimize travel (e.g., bus routing) alongside integration---or even algorithms that select from a menu of different integration strategies (e.g., redistricting or pairings/triplings) for a particular district---may help produce new policy possibilities that offer practical pathways to more integrated schools.

In summary, across many school districts, elementary school mergers can foster more racially/ethnically integrated schools without imposing large travel burdens on families. We hope this preliminary investigation helps equip district leaders and communities with additional evidence and tools to support their efforts to foster more inclusive learning environments.

\section{Materials and Methods}

\subsection{Data sample}
We begin with 2021/2022 school enrollments by race/ethnicity and grade level from the US Department of Education's Common Core of Data, along with Census block-to-school mappings released by~\citep{redrawing} via elementary school attendance boundaries purchased for the 2021/2022 school year.  We narrow to the 6,228 local education agencies (``districts'') with at least four elementary schools---i.e., districts where there are at least a few different ways elementary schools might be merged. The data released by~\citep{redrawing} filtered out elementary schools that permit out-of-boundary attendance (like magnet programs whose attendance boundary overlaps with the entire district); therefore, these schools are not included in our simulations and accompanying analyses. This is appropriate for our study given that many open-enrollment schools may have specialty programs or themes (Arts, STEM, etc.) that could make it impractical to merge them with other schools not offering such programs. As in~\citep{redrawing}, it is also impossible to estimate travel times for out-of-boundary attendees at open enrollment schools. 

After filtering out these open enrollment schools, we select the top 200 districts (by closed enrollment elementary school population) to include in our analysis (with a few large districts---namely, Los Angeles Unified, Chicago Public Schools, and Houston Independent School District not included because the memory requirements for simulating changes in them are too demanding). Across these 200 districts, there are 9,007 elementary schools deemed as non-open-enrollment, which serve nearly 4.5 million elementary students. Our selection of 200 as the number to focus on in this study is somewhat arbitrary and driven by an effort to focus the analysis on a limited, yet still heterogeneous, sample. As our results indicate, results can vary across different districts, which further motivates the release of the online dashboard described earlier and inclusion in that dashboard of simulation results for thousands of additional districts.

Importantly, we note that the original boundaries data used in~\citep{redrawing} was purchased by a commercial data provider that determined a school's open enrollment status by simply observing if the school's boundaries overlapped with any other attendance boundaries in the district (if so, the school is deemed as allowing open enrollment). This means that within district choice, transfer, and other programs that do not expand the attendance boundary of a school to make them overlap with others, may not be accurately captured by that provider---and hence, it is possible that some of the schools included in this study may have enrollments by students living outside of the school's designated attendance boundaries. We note this as a limitation of our data and approach that can be corrected in the future through collaborations with specific districts (who have more accurate ground truth on school enrollments), and/or an updated attendance boundary survey by the Department of Education. 

We measure segregation using the dissimilarity index~\citep{massey1988dimensions}. Dissimilarity is a popular measure of segregation, though it suffers from a number of limitations~\citep{james1985seg,winship1978dissim}. We use it here given its simplicity and the fact that prior work has found alignment between the results of simulated (redistricting) integration strategies that optimize for dissimilarity and those optimizing for other measures, like the gini and variance ratio indices~\citep{redrawing}.  

In comparison to the total set of 6,228 districts, the 200 districts in our sample have a much smaller median percentage of total elementary students who are white (approx. 32\% versus 76\%)---and they are much more segregated (median dissimilarity index of 0.4 vs. 0.2 in the full set of districts).  

\subsection{Optimization model}

We design a school mergers algorithm based on the datasets described above.  The algorithm's goal is to determine (i) an assignment of schools to other schools (i.e. school mergers), and (ii) which grades each school should subsequently serve in order to minimize dissimilarity across the district, subject to a number of constraints.  These constraints include allowing only adjacent schools to be paired or tripled; requiring schools to serve contiguous, non-overlapping grade spans; and ensuring that both the current and projected future enrollments of a school (i.e., if it serves an older grade span in a pairing/triple) are within school capacity constraints.  Given that this assignment problem is NP-hard, we use the CP-SAT constraint programming solver in Google's OR tools library \citep{ortools} to efficiently identify solutions.

As data inputs, the algorithm accepts a set of schools that can be merged (i.e., boundary-adjacent schools), and the student enrollments per grade, per race/ethnicity at each of those schools. Unlike redistricting~\cite{redrawing}, individual Census blocks/student populations in individual blocks are not inputs. Block-level student populations are estimated only to estimate impacts on driving times, but also unlike prior redistricting work, these driving times are not inputs into the mergers algorithm.

For each district, we define two binary matrices.  The first is $M^{|S|\times|S|}$, where $S$ is the set of elementary schools in the district.  $M_{ij}=1$ implies that schools $i$ and $j$ should be merged. A school is always considered as ``merged'' with itself, i.e., $M_{ii} = 1$ for all $i$.  The second is $R^{|S|\times|G|}$, where $G$ is the set of grades served across $S$.  $R_{ij}=1$ implies that school $i$ serves grade level $j$.  The entries of these two matrices represent the key decision variables for the algorithm.

Our primary objective is to minimize each $d$'s ``dissimilarity'', defined as the dissimilarity index of segregation across White students and students of color, i.e.:

\begin{equation}
    \label{eq:1}
    D = \frac{1}{2} \sum_{s \in S} \left|\frac{w_s}{w_T} - \frac{t_s - w_s}{T - w_T}\right|
\end{equation}

Here, $s$ is an elementary school across all district elementary schools $S$; $t_s$ and $w_s$ indicate the total and total White students at $s$, respectively; and $T$ and $w_T$ indicate the total and total White students across the district, respectively. Crucially, $w_s$ and $t_s$ depend on the values of $M_{s,:}$ and $R_{s,:}$.

Prior work has shown that optimizing for dissimilarity as the measure of segregation when changing student assignment policies produces similar results to optimizing for some other measure of segregation (like the variance ratio index)~\citep{redrawing}. Diversifying schools is a multi-dimensional challenge, and dissimilarity, while limited, provides a clear \& consequential metric for progress. Future work may focus more granularly on the ethnoracial dynamics of school mergers by using diversity measures like the $M$ measure~\citep{elbers2021method,frankenberg2023demo} that can be decomposed into the sources of change, for example, unit-margin changes due to school sizes scaling up/down with racial group proportions staying the same, group-margin changes due to race/ethnic groups entering the system (relevant to the interdistrict simulations described in Section~\ref{sec:sensitivity_interdistrict}), or other more structural changes.

We impose the following constraints.  First, we require that schools can only be paired, tripled, or left unchanged:

\begin{equation}
    \label{constraint:pair_triple}
    \sum_{s^\prime \in S} M_{s,s^\prime} \in \{1,2,3\}
    \qquad \forall s \in S
\end{equation}

Next, we require that mergers are symmetric and transitive:

\begin{equation}
    \label{constraint:symmetric}
    M_{s,s^\prime} = 1 \implies M_{s^\prime,s} = 1
    \qquad \forall s, s^\prime \in S
\end{equation}

\begin{equation}
    \label{constraint:transitive}
    M_{s,s^\prime} = 1 \land ~M_{s^\prime, s^{\prime\prime}} = 1 \implies M_{s,s^{\prime\prime}} = 1
    \qquad
    \forall s, s^\prime, s^{\prime\prime} \in S
\end{equation}

Next, we require that each grade level is assigned to exactly one school in a given merger: 

\begin{equation}
    \label{constraint:pair_triple}
    M_{s,s^\prime} = 1 \implies ~R_{s,g} + R_{s^\prime, g} = 1
    \qquad
    \begin{matrix}
    \forall s \in S,\ \forall s^\prime \in S \setminus \{s\},\\
    \forall g \in G
    \end{matrix}
\end{equation}

Next, the grade span served by any particular school $s \in S$ in a merger must be contiguous:

\begin{equation}
    \label{constraint:contig}
    R_{s,g} =
    \begin{cases} 
    1, & \mathrm{if}\ g_s^{\mathrm{start}} \leq g \leq g_s^{\mathrm{end}} \\
    0, & \mathrm{otherwise} 
    \end{cases}
    \qquad \forall s \in S,\ \forall g \in G
\end{equation}

\noindent
where $g_s^{\mathrm{start}}$ and $g_s^{\mathrm{end}}$ are the minimum and maximum grades served by school $s$.

Next, each school's resulting enrollment must be within a pre-specified minimum and maximum capacity:

\begin{equation}
    \label{constraint:cap_current}
    p_\mathrm{min} \cdot \sum_{g \in G} ~E_{s,g} \leq \sum_{g \in G} ~\sum_{s^\prime \in S} ~M_{s,s^\prime} \cdot R_{s,g} \cdot E_{s^\prime,g} \leq \mathrm{Capacity}(s)
    \qquad \forall s \in S
\end{equation}

\noindent
Here, $E_{s,g}$ indicates how many students in grade $g$ are currently enrolled at school $s$; $p_\mathrm{min}$ is a value between 0 and 1 representing the minimum fraction of a school's current population size that must be enrolled at the school post-merger (we set this to 0.8); and $\mathrm{Capacity}(s)$ indicates the school's maximum capacity.  To our knowledge, the Department of Education does not report school capacities, likely in part because the capacity of a school can be flexible (i.e. if schools allow the addition of trailers for classrooms; classrooms that can accommodate multiple classes at different times; etc).  In the absence of ground truth data on school capacities, we (crudely) estimate them as the maximum enrollment observed at the school from 2016/2017 through 2021/2022.  

Each school's future enrollment (i.e., the enrollment expected at a merged school that serves a higher grade span than other schools it has been merged with, once the students from lower grades move up to these higher grades) must also be within the same minimum and maximum capacity.  That is, for any two schools $s$ and $s^\prime$:

\begin{equation}
    \label{constraint:later_cap}
    g_{s^\prime}^{\mathrm{end}} > g_s^{\mathrm{end}} \land M_{s,s^\prime} = 1 \implies
    p_\mathrm{min} \cdot \sum_{g \in G} ~E_{s^\prime,g} \leq \sum_{g \in G} ~\sum_{s^{\prime\prime} \in S} ~M_{s,s^{\prime\prime}} \cdot R_{s,g} \cdot E_{s^{\prime\prime},g} \leq \mathrm{Capacity}(s^\prime)
\end{equation}

Finally, only schools with adjacent attendance boundaries are eligible to be merged:

\begin{equation}
    \label{constraint:adjacent}
    M_{s,s^\prime} = 1 \implies \mathrm{Adjacent}(s,s^\prime)
    \qquad \forall s,s^\prime \in S
\end{equation}

Where $\mathrm{Adjacent}(s,s^\prime)$ is a predicate that is true if and only if schools $s$ and $s^\prime$ have attendance boundaries that touch.

\subsection{Simulations}

We implement the above model using the CP-SAT constraint programming solver in Google's OR tools library \citep{ortools} to efficiently identify solutions.  Simulations are run using a single CPU core on a parallel computing cluster, with a maximum runtime of 5.5 hours.  The algorithm produces the optimal solution for 19\% (38 out of 200) districts; for the remaining 81\% (162 out of 200), it produces a feasible solution but is not able to prove that the solution is optimal under the specified objective and constraints.

\subsection{Estimating travel times}

We estimate travel times using the OpenRouteService API~\citep{ors2024}, compiled and run on a server using an Open Street Map of the US downloaded in the Fall of 2021.  Like in~\citep{redrawing}, travel times are represented as estimated driving times from the centroid of a given census block to a particular school.  Importantly, to our knowledge, the OpenRouteService API does not factor in traffic patterns (similar to~\citep{redrawing}).  Estimating a merger's potential impact on travel times requires estimating how many students from each racial/ethnic group, and in each grade level, live in each block and attend the correspondingly-assigned school.  We use the method from~\citep{redrawing} to estimate the total number of students from each racial/ethnic group who live in each block, irrespective of grade level: essentially, this method uses Census block-level estimates of number of non-Adults, per race/ethnicity, living in each Census block in order to estimate the fraction of a school's enrollment belonging to that racial/ethnic group that can be attributed to living in those blocks.  We are unable to estimate grade-level counts because the Census data does not have this level of granularity available.  Therefore, we assume that the students of a particular racial/ethnic group involved in a school merger are distributed across Census blocks at the same rate as their overall population at a particular school. 

The following example may help explain this method: imagine there are 4 Census blocks assigned to attend ABC elementary which serves grades K-5, and there are 100 Black students living across these 4 blocks (each of which contains 30, 25, 15, 40 Black students, respectively). Now, imagine ABC elementary pairs with XYZ elementary (also serving grades K-5), and ABC is now set to serve grades K-2, while XYZ will serve grades 3-5 post-pairing.  If there are 40 Black students in grades 3-5 who were originally assigned to ABC and will now attend XYZ post-pairing, we assume that (30/100) * 40 live in the first block, (25/100) * 40 live in the second, etc.---i.e., that the percentage of students from a particular racial/ethnic group across the grades living in each block and involved in a merger which would move them from some status quo school to some other school is equivalent to the overall percentage of students from a particular racial/ethnic group living in that block and attending the status quo school. This estimation procedure may yield fractional students, which we accept, given it does not impede our main objective of approximating average impacts that mergers might have on travel times. We simply then multiply this (possibly fractional) number of school switchers per racial/ethnic group, per block by the travel time to the new school to estimate the new driving times for those students. We would perform a similar computation for the K-2 graders who would be re-assigned from school XYZ to school ABC.

A key assumption underlying this method is that the fraction of students in a given grade and living in a given block that are from a particular racial/ethnic group is equal to the fraction of students in a particular grade at a school overall that are from that racial/ethnic group. This may not hold if there are imbalances in how populations are distributed within an attendance boundary (e.g. if some parts of a boundary tend to have higher concentrations of White first graders than others), and would be an important assumption to further explore and test with ground-truth data from districts gathered through researcher-practitioner partnerships.

\subsection{Computing spatial autocorrelation}

To calculate Geary's $C$, we construct an adjacency matrix $\mathrm{W}$, where, for adjacent schools $s$ and $s^\prime$, $\mathrm{W}_{s,s^\prime}$ is the total population of $s$.  Non-adjacent schools or schools connecting with themselves have an edge weight of 0. Schools without neighbors are excluded from $\mathrm{W}$, and $\mathrm{W}$ is row standardized to ensure validity and interpretability. Then, Geary's $C$ is given by

\begin{equation}
C = \left(
   \sum_{s\in S} \sum_{s^\prime\in S} \mathrm{W}_{s,s^\prime} (x_s-x_{s^\prime})^2
\right)
\left(
   \frac{2}{|S| - 1} \sum_{s\in S} (x_s-\bar{x})^2 \sum_{s\in S} \sum_{s^\prime\in S} \mathrm{W}_{s,s^\prime}
\right)^{-1}
\end{equation}

\noindent where $\bar{x}$ is the proportion of White students across the district.

\section{Acknowledgments}
We are grateful to Alvin Chang, Halley Potter, Genevieve Siegel-Hawley, Jonathan Sotsky, School Board members in New York City's District 13, and students in Northeastern University's Interdisciplinary Design and Media PhD program for their thoughtful input to help shape and improve this project. We are also grateful to the Anonymous Reviewers of our manuscript for their helpful suggestions.

\section{Supplementary Materials}
Supplementary materials can be accessed here: \url{https://drive.google.com/file/d/1nolX5mPCgBoQH6YMn7QAOtebfWljPdyu/view}.

\section{Funding}
This work is supported by funding from the Overdeck Family Foundation.

\section{Author contributions statement}
M. L. and N. G. both contributed to designing the project, conducting analyses, and writing.

\section{Conflicts of interest}
Editor Esteban Moro and authors Madison Landry and Nabeel Gillani are affiliated with Northeastern University. They have not collaborated on this or other studies. The authors declare no other competing interests.

\section{Data availability}
All code and data required to replicate the main and supplementary results of the paper can be found here: \url{https://github.com/Plural-Connections/public-school-mergers}.

\bibliographystyle{apacite}
\bibliography{main_v2}

\end{document}